\documentclass[preprint,12pt]{elsarticle}
\usepackage{amssymb}
\usepackage{amsmath}
\usepackage{amsthm}

\usepackage{graphicx}
\usepackage{subcaption}
\usepackage{tabularx}

\journal{Computer}

\begin{document}

\begin{frontmatter}


\title{Deep Reinforcement Learning for Fault-Adaptive Routing in Eisenstein-Jacobi Interconnection Topologies}

\author[1]{Mohammad Walid Charrwi}
\author[1]{Zaid Hussain}

\affiliation[1]{organization={High Performance Computing Lab, Computer Science Department, Kuwait University},country={Kuwait}}

\begin{abstract}
The increasing density of many-core architectures necessitates interconnection networks that are both high-performance and fault-resilient. Eisenstein-Jacobi (EJ) networks, with their symmetric 6-regular topology, offer superior topological properties but challenge traditional routing heuristics under fault conditions. This paper evaluates three routing paradigms in faulty EJ environments: deterministic Greedy Adaptive Routing, theoretically optimal Dijkstra's algorithm, and a reinforcement learning (RL)-based approach. Using a multi-objective reward function to penalize fault proximity and reward path efficiency, the RL agent learns to navigate around clustered failures that typically induce dead-ends in greedy geometric routing. Dijkstra's algorithm establishes the theoretical performance ceiling by computing globally optimal paths with complete topology knowledge, revealing the true connectivity limits of faulty networks. Quantitative analysis at nine faulty nodes shows greedy routing catastrophically degrades to 10\% effective reachability and packet delivery, while Dijkstra proves 52-54\% represents the topological optimum. The RL agent achieves 94\% effective reachability and 91\% packet delivery making it suitable for distributed deployment. Furthermore, throughput evaluations demonstrate that RL sustains > 90\% normalized throughput across all loads, actually outperforming Dijkstra under congestion through implicit load-balancing strategies. These results establish RL-based adaptive policies as a practical solution that bridges the gap between greedy's efficiency and Dijkstra's optimality, providing robust, self-healing communication in fault-prone interconnection networks without requiring the global topology knowledge or computational overhead of optimal algorithms.

\end{abstract}

\begin{keyword}
Distributed Computing \sep Faulty Nodes \sep Greedy Adaptive \sep Hexagonal Networks \sep Routing \sep Reinforcement Learning.
\end{keyword}

\end{frontmatter}

\section{Introduction}
\label{section:introduction}

The continuous advancement of multi-core and many-core architectures has positioned the interconnection network as a primary determinant of overall system performance, energy efficiency, and reliability \cite{hussain2016improvedonetoallbroadcastinghigher}. Modern many-core systems and High-Performance Computing (HPC) architectures rely on efficient Network-on-Chip (NoCs) to interconnect a growing number of processing elements. As the number of processing elements integrated into a single Very Large Scale Integration (VLSI) chip scales toward the thousand-core threshold, the limitations of traditional 2D mesh and torus topologies characterized by their relatively high diameter and limited connectivity become apparent \cite{RLNOC}. In the pursuit of more efficient communication fabrics, researchers have turned to quadratic residue-based networks, specifically Gaussian and Eisenstein-Jacobi (EJ) networks \cite{BoseEJ}. Eisenstein–Jacobi (EJ) networks offer a compelling alternative by forming a hexagonally connected toroidal NoC derived from complex Eisenstein integers. These EJ networks have $|\alpha|^2 = a^2 + ab + b^2$ nodes generated by an Eisenstein integer $\alpha = a + b\omega$, with wrap-around links defined by the congruence classes mod $\alpha$. EJ topologies possess higher node connectivity and a lower diameter compared to rectangular grids of equivalent size, significantly enhancing communication efficiency. Consequently, hexagonal torus NoCs have emerged as attractive candidates for high-bandwidth, low-latency interconnects \cite{EJBroadcast}. 

However, the performance of these complex topologies is highly sensitive to link or node faults, which can severely degrade network reliability. Traditional minimal greedy routing, which operates by forwarding packets to the neighbor that minimizes Euclidean distance to the destination, is computationally simple but suffers from inherent brittleness. In the presence of failures, greedy algorithms frequently become "trapped" in local minima, where no available neighbor provides a reduction in distance, leading to catastrophic delivery failures and excessive detours \cite{charrwi2025resilientpacketforwardingreinforcement}. At the opposite extreme, globally optimal routing algorithms such as Dijkstra's shortest path can compute theoretically optimal routes by exhaustively exploring network topology with complete fault knowledge. While Dijkstra establishes the performance ceiling—guaranteeing shortest paths whenever they exist—its $O(|V|\log|V|)$ computational complexity and requirement for global topology state make it impractical for per-packet routing decisions in large-scale distributed NoC systems.

By contrast, reinforcement learning (RL)–based routing has the capacity to learn global topological features and dynamically navigate around faulty regions while maintaining the computational efficiency and local decision-making characteristics suitable for distributed deployment. Recent NoC studies demonstrate this potential. For instance, the DeepNR algorithm utilizes a deep RL agent to achieve 21–44\% lower latency and > 90\% throughput under high load compared to traditional heuristics \cite{DeepNR}. Furthermore, research in \cite{charrwi2025selfhealingnetworksonchiprldrivenrouting} demonstrates that a Proximal Policy Optimization (PPO)-based RL router in a 2D torus maintains a packet delivery ratio (PDR) > 90\% even under significant fault densities, whereas minimal adaptive schemes experience a sharp collapse to ~70\%. 

In this work, we evaluate RL-driven routing against both a greedy baseline and Dijkstra's optimal algorithm within the context of an EJ NoC subjected to node failures. We employ a PPO-based training regime where each router acts as an independent agent, choosing among six EJ neighbors to forward packets. The reward function is designed to heavily penalize transitions into faulty nodes while mildly penalizing additional hops to encourage efficiency. The greedy baseline always forwards to the geometrically closest live neighbor, dropping packets when progress is impossible. Dijkstra computes globally optimal shortest paths with complete topology knowledge, establishing the theoretical performance ceiling. Our experimental evaluation under uniform random traffic reveals that the RL policy dramatically outperforms greedy routing while approaching Dijkstra-level performance across all fault metrics. Specifically, at nine faulty nodes, the RL agent maintains 94\% effective reachability and 91\% packet delivery ratio compared to greedy's catastrophic 10\% for both metrics, while Dijkstra establishes that 52\% reachability and 54\% delivery represent the topological optimum. The RL approach achieves near-optimal performance (within 42 percentage points of Dijkstra's ceiling ) while operating with only local information and efficient per-packet decision-making, demonstrating that learning-based methods can bridge the gap between greedy's computational efficiency and Dijkstra's theoretical optimality.

This paper is organized as follows. The related work on routing with faults is discussed in Section \ref{section:relatedwork}. In Section \ref{section:background}, we describe the graphical and mathematical concepts related to Eisenstein-Jacobi interconnected networks. The methodology for mitigating faults and the RL formulation for routing is explained in Section \ref{section:methadology}. Next in Section \ref{section:experimental}, we present simulation results and network metrics that highlight the advantages of utilizing an RL-based approach over greedy routing and its proximity to Dijkstra's theoretical optimum in the presence of faults. Finally, the paper is concluded in Section \ref{section:conclusion} highlighting the important takeaways from the proposed work and discussion of future directions.

\section{Related Work}
\label{section:relatedwork}
The design of interconnection networks has historically balanced the trade-offs between degree, diameter, and wiring complexity \cite{KingNetwork}. While the 2D mesh remains a staple due to its planar layout compatibility, its degree-four limitation results in a diameter that grows linearly with the side length, creating bottlenecks in large-scale systems. To mitigate this, the Eisenstein-Jacobi network was proposed as a 6-regular symmetric topology that generalizes the hexagonal mesh. The EJ network stems from the quotient ring of Eisenstein-Jacobi integers. Unlike standard integers, it represents a triangular lattice in the complex plane, where each point is equidistant from its six immediate neighbor \cite{hussain2016improvedonetoallbroadcastinghigher}.

\subsection{Fault-Adaptive Routing}
The quest for reliable communication in faulty networks has produced a vast body of literature, ranging from static protection schemes to highly adaptive dynamic protocols. Traditional routing and geometric forwarding strategies have long struggled with scalability and fault resilience \cite{greedyRoutingFailure}. Geometric routing, which relies on nodes' coordinates rather than global lookup tables, is particularly susceptible to local minima in the case where all functional neighbors are farther from the destination than the current node \cite{GreedyPostitionFailure}. Another approach is fault detection mechanism include link-state advertisements or explicit failure notifications. The Open Shortest Path First (OSPF) protocol exemplifies this approach, recomputing routing tables upon topology changes. However, such methods suffer from convergence delays during which routing may be inconsistent or suboptimal. Recent work by \cite{kushman2007rbgp} on source routing with failure recovery has attempted to embedding backup path information within packet headers. 

\subsection{Reinforcement Learning in Interconnection Networks}
The application of reinforcement learning to network routing has evolved significantly since the pioneering work of Q-learning in \cite{boyan1994packet} which modeled routing decisions as a Markov Decision Process (MDP) where routers learn to estimate delivery times to destinations and select next-hop neighbors that minimize these estimates. Early RL applications in networking utilized tabular methods like Q-Learning or simple task migration or routing in Software Defined Networks (SDN) \cite{RLSDN}. While these methods demonstrated the ability to balance load and avoid congestion, they struggled with the large state spaces inherent in high-performance interconnection networks. The integration of Reinforcement Learning (RL) into network routing represents a paradigm shift from hand-crafted heuristics to learned, self-adaptive policies \cite{charrwi2025resilientpacketforwardingreinforcement}. The advent of Deep Reinforcement Learning (DRL) and policy gradient methods, such as Proximal Policy Optimization (PPO), has enabled agents to handle high-dimensional state representations (such as those found in Gaussian networks) while maintaining training stability \cite{charrwi2025selfhealingnetworksonchiprldrivenrouting}. PPO (Proximal Policy Optimization) algorithm, in particular, has emerged as a state-of-the-art solution for NoC and HPC routing due to its clipped surrogate objective, which prevents large, destabilizing updates during the learning process. The utilization of RL have advanced different fields including fields like hardware design \cite{mirhoseini2021chip}, quantum computing \cite{charrwi2025tpu, rlquantum} and many other fields. Work has have been successfully applied PPO-based RL to bypass faulty regions in Gaussian networks \cite{charrwi2025resilientpacketforwardingreinforcement}, demonstrating a Packet Delivery Ratio (PDR) of 0.95 at 40\% fault density, a significant improvement over the 0.66 PDR achieved by greedy adaptive routing. This research builds upon those foundations by applying the PPO-based RL framework to the more complex EJ topology, examining how the 6-regular structure influences the agent's ability to learn detour strategies in the presence of various fault models.

\section{Background}
\label{section:background}
The design of interconnection networks has historically balanced the trade-offs between degree, diameter, and wiring complexity. While the 2D mesh remains a staple due to its planar layout compatibility, its degree-four limitation results in a diameter that grows linearly with the side length, creating bottlenecks in large-scale systems. To mitigate this, the Eisenstein-Jacobi network was proposed as a 6-regular symmetric topology that generalizes the hexagonal mesh. The mathematical elegance of the EJ network stems from the quotient ring of Eisenstein-Jacobi integers, denoted as $\mathbb{Z}[\rho]$, where $\rho = e^{i\pi/3} = \frac{1 + i\sqrt{3}}{2}$. Unlike standard integers, $\mathbb{Z}[\rho]$ represents a triangular lattice in the complex plane, where each point is equidistant from its six immediate neighbors. The total number of nodes in an EJ network, governed by a generator $\alpha = a + b\rho$, is defined by the norm $N(\alpha) = a^2 + b^2 + ab$. This formula highlights a fundamental advantage: for a given side length $a$, the EJ network supports a substantially higher node count than a square mesh ($a^2$) while maintaining a smaller diameter. For example, a Gaussian network with 200 nodes might possess a diameter of 10, whereas a standard 2D torus of the same size would have a diameter of at least 15. The EJ network further optimizes this by providing six unit directions $\{ \pm 1, \pm \rho, \pm \rho^2 \}$ for packet forwarding, compared to the four directions in Gaussian or mesh topologies. This inherent redundancy is not merely a topological curiosity; it provides the mathematical basis for decomposing the network into three edge-disjoint Hamiltonian cycles, offering multiple non-overlapping paths for fault recovery and secure broadcasting \cite{BoseEJ}. In a formal representation, the EJ network is modeled as a graph $EJ_\alpha = (V, E)$, where the vertex set $V$ corresponds to the residue classes of $\mathbb{Z}[\rho]$ modulo $\alpha$. The distance between two nodes, $\beta$ and $\gamma$, is not computed via standard Euclidean metrics but through the $\rho$-taxicab norm, which reflects the number of hops across the 6-regular grid. The distance $D_\alpha(\beta, \gamma)$ is given by:

\[
\begin{aligned}
D_\alpha(\beta, \gamma) = \min \{ & |x| + |y| + |z| \mid (\beta - \gamma) \equiv x + y\rho + z\rho^2 \pmod \alpha \}
\end{aligned} 
\]

The simplicity of this distance calculation allows for efficient shortest-path routing in healthy networks. The diameter of the network, $M$, is bounded by the relation $M = \lfloor (a+2b)/3 \rfloor$ for $0 \le a \le b$, ensuring that the maximum latency between any two processors is strictly controlled. Furthermore, the distribution of distances follows a predictable hexagonal expansion, where the number of nodes at distance $t$ is $6t$ for $1 \le t < (a+b)/2$, effectively providing a wide selection of routing options for adaptive protocols. This structural richness is the catalyst for exploring advanced RL-based routing, as it provides the state space complexity required for an agent to learn non-trivial, fault-aware navigation strategies.

\begin{figure}[t!]
    \centering
    \includegraphics[width=0.9\linewidth]{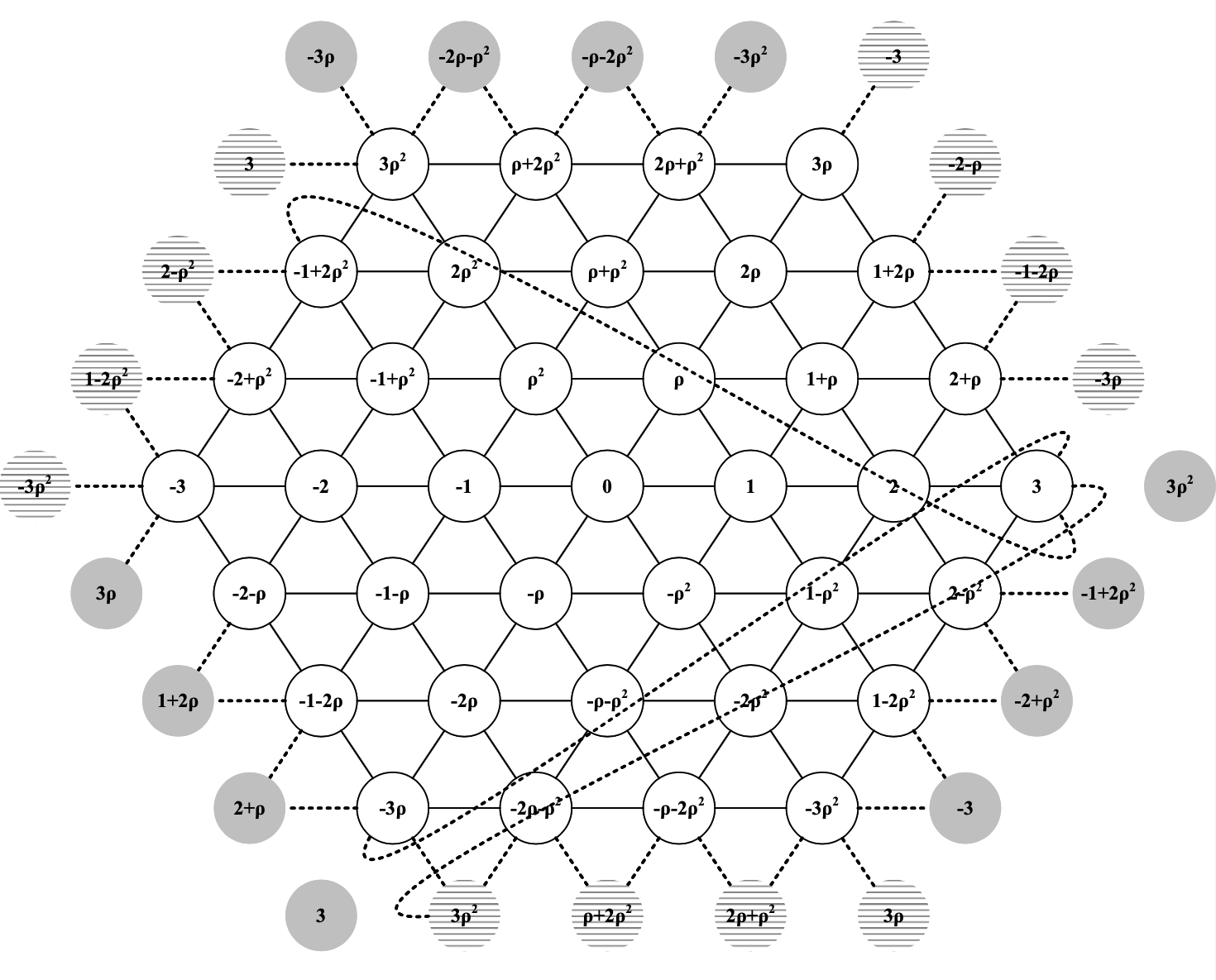}
    \caption{EJ network generated with $\alpha = 3 + 4\rho$} 
    \label{fig:EJNetwork}
\end{figure}

\section{Methodology}
\label{section:methadology}
The comparative evaluation is framed as a Markov Decision Process (MDP) for the RL agent, while the greedy adaptive baseline follows a deterministic, local optimization heuristic. The EJ network environment is simulated by mapping the residue classes modulo $\alpha$ to a 2D state space where each node is aware of its coordinates and the status of its six immediate neighbors.

\subsection{Fault Model}
The experimental setup subjects both protocols to two types of stress tests: increasing fault density and increasing network load. Faults are injected by randomly disabling a percentage of nodes. 
To simulate thermal hotspots, a \textit{clustered fault model} is also used, where failures are concentrated in specific regions, creating a local minima region that are most challenging for greedy routing. Traffic load is modeled by varying the "offered load" (the rate at which new packets are injected into the network) to measure the impact of congestion and buffer management on routing success

\subsection{Greedy Adaptive Routing}
The greedy adaptive protocol serves as the control, representing the current standard for low-overhead routing in structured topologies. At each hop, the algorithm calculates the distance from all operational neighbors to the destination using the Euclidean norm in the complex plane: $dist = |z_{next} - z_{dest}|$. The neighbor that provides the maximum reduction in distance is selected for the next hop. If the primary greedy neighbor is faulty, the algorithm adaptively selects the next best functional neighbor. Crucially, if no functional neighbor improves the distance toward the destination (for instance, the packet is in a local minimum), the packet is dropped, contributing to routing failure. This protocol is "memory-less" and "stateless," relying only on the immediate local neighborhood. For comparison, we implement a greedy routing approach that at each hop, selects the neighboring node that minimizes the distance to the destination. In failure-free Eisenstein--Jacobi networks, greedy routing is shortest-path
optimal, as each forwarding step strictly decreases the Eisenstein distance to
the destination, consistent with distance-monotone routing in regular lattice
and mesh interconnection networks \cite{duato2003interconnection}. However, greedy routing lacks any mechanism for exploration when the locally optimal neighbor is faulty or leads to a dead end. Consequently, greedy routing fails whenever it encounters a local minimum in which all available neighbors are either faulty or farther from the destination than the current node.





\subsection{Reinforcement Learning Routing}
The routing task for the RL agent is defined by the tuple $(S, A, P, R, \gamma)$, where $S$ is the state space, $A$ is the action space, $P$ is the transition probability, $R$ is the reward function, and $\gamma$ is the discount factor. 

\begin{itemize}
    \item State Space ($S$): Each state $s_t$ encompasses the current node's EJ coordinates $(x, y)$, the destination ($d$) coordinates $(x_d, y_d)$, and a local connectivity mask indicating which of the six neighbors are operational.
    \item Action Space ($A$): The agent selects one of six discrete actions, corresponding to the unit directions $\{ \pm 1, \pm \rho, \pm \rho^2 \}$.
    \item Transition Model ($P$): Transitions are deterministic; selecting an action at a give timestep $t$ is $a_t$ at node $s_t$ moves the packet to node $s_{t+1}$ if the node is functional. If the node is faulty or out of bounds, the episode terminates.
    \item Reward Function ($R$): To drive both efficiency and resilience, a multi-objective reward function is employed. Reaching the destination yields a significant positive reward ($r = +100$), while colliding with a faulty node incurs a severe penalty ($r = -50$). To encourage shortest-path behavior, a continuous step cost is applied for every hop ($r = 1$).
    \item Optimization via PPO-based RL: The agent utilizes an actor-critic architecture, where the actor network determines the policy $\pi_\theta(a|s)$ which maps a given network state $s$ to a probability distribution over the available actions $a$, parameterized by the neural network weights $\theta$. Whereas the critic network estimates the value function $V_\phi(s)$, which estimates the expected cumulative reward the agent will receive starting from state $s$. Parameterized by weights $\phi$, the critic provides a baseline estimate used to calculate the advantage of an action. The use of Generalized Advantage Estimation (GAE) helps reduce variance in policy updates, while the clipping parameter $\epsilon$ (set to 0.2) ensures that the updated policy does not deviate excessively from the previous iteration, maintaining stability in the volatile, fault-prone environment.
\end{itemize}

\subsection{Dijkstra's Shortest Path Routing}
Dijkstra's algorithm serves as the theoretical optimality baseline, representing the the well-known achievable routing performance with complete global topology knowledge. The algorithm computes globally shortest paths by systematically exploring the network from the source, maintaining a priority queue of nodes ordered by distance. At each iteration, it selects the unvisited node with minimum distance, examines operational neighbors, and updates distances if shorter paths are found. In faulty EJ networks, Dijkstra operates with complete real-time knowledge of all fault locations, treating faulty nodes as absent from the graph. The algorithm identifies routes requiring temporary distance increases to circumnavigate faults—detours that greedy algorithms refuse. Dijkstra guarantees finding the shortest path when one exists and correctly identifies topologically disconnected pairs. The computational complexity is $O(|E| + |V|\log|V|)$, or $O(|V|\log|V|)$ in 6-regular EJ networks where $|E| = 3|V|$. This overhead becomes prohibitive for per-packet routing in high-throughput networks as size scales. Additionally, Dijkstra requires maintaining complete global topology state and disseminating real-time fault information network-wide, creating substantial communication and storage overhead. Despite these practical limitations, Dijkstra's performance establishes ground truth for our evaluation. By comparing greedy and RL against Dijkstra's optimal paths, we quantify performance loss from local decision-making and limited information. Dijkstra's effective reachability reveals true network connectivity, distinguishing routing algorithm failures from fundamental network partitioning, while its path lengths provide the optimal baseline for measuring routing efficiency.
$\\$

We evaluate routing performance under varying fault densities and traffic loads using three metrics:
\begin{itemize}
    \item \textbf{Packet Delivery Ratio (PDR):} The fraction of packets successfully delivered to their destination (i.e. not trapped or dropped). It is computed as:

    \[
    \text{PDR}= \frac{\text{Total Packets Received}} {\text{Total Packets Sent}}
    \]
    
    \item \textbf{Effective Reachability Ratio (ERR):} The fraction of source--destination pairs for which the routing algorithm successfully delivers a packet. It is computed as:
    \[
    \text{ERR}= \frac{\text{Number of successful deliveries}} {\text{Total routing instances}}
    \]
    
    \item \textbf{Normalized Throughput:} Effective delivery rate under congestion, modeled as $(\text{packets delivered}) / (\text{injection rate})$, essentially penalized by both path length and queuing. A throughput model with inverse-exponential dependence on path length is used, so shorter paths significantly boost throughput.
\end{itemize}

The RL agent operates using an Actor-Critic architecture, where the Policy Network ($\pi_{\theta}$) learns a stochastic mapping from state to action probabilities, and the Value Network ($V_{\phi}$) estimates the expected return $V(s)$. Training is conducted on-policy, collecting a batch of trajectories $\mathcal{D}$ before updating both networks multiple times using the Generalized Advantage Estimation (GAE) and the Clipped Surrogate Objective to stabilize gradient updates. The core of the PPO-based RL optimization involves the following steps: Advantage Calculation (GAE): The advantage $\hat{A}_t$ at time $t$ is calculated using the Value Network's temporal difference error $\delta_t$ at time $t$ to reduce variance:
\begin{equation}
\delta_t = R_t + \gamma V_{\phi}(s_{t+1}) - V_{\phi}(s_t)
\end{equation}

\begin{equation}
\hat{A}_t = \sum_{l=0}^{T-t} (\gamma\lambda)^l \delta_{t+l}
\end{equation}

where $\gamma=0.95$ is the discount factor and $\lambda=0.92$ is the GAE parameter. Policy Update (Clipped Objective): The Policy Network is updated by maximizing the PPO-based RL objective function, which constrains the probability ratio $r_t(\theta)$ between the new policy ($\pi_{\theta}$) and the old policy ($\pi_{\theta_{\text{old}}}$) within a clipping margin $\epsilon=0.2$:$$L^{\text{CLIP}}(\theta) = \hat{\mathbb{E}}_t \left[ \min\left( r_t(\theta) \hat{A}_t, \text{clip}(r_t(\theta), 1-\epsilon, 1+\epsilon) \hat{A}_t \right) \right]$$where $r_t(\theta) = \frac{\pi_{\theta}(a_t|s_t)}{\pi_{\theta_{\text{old}}}(a_t|s_t)}$. The Value Network is simultaneously updated by minimizing the Mean Squared Error (MSE) loss: 
\begin{equation}
L^{\text{VF}}(\phi) = \hat{\mathbb{E}}_t \left[ (R_t - V_{\phi}(s_t))^2 \right]
\end{equation}

\begin{figure}[t!]
    \centering
    \includegraphics[width=0.7\linewidth]{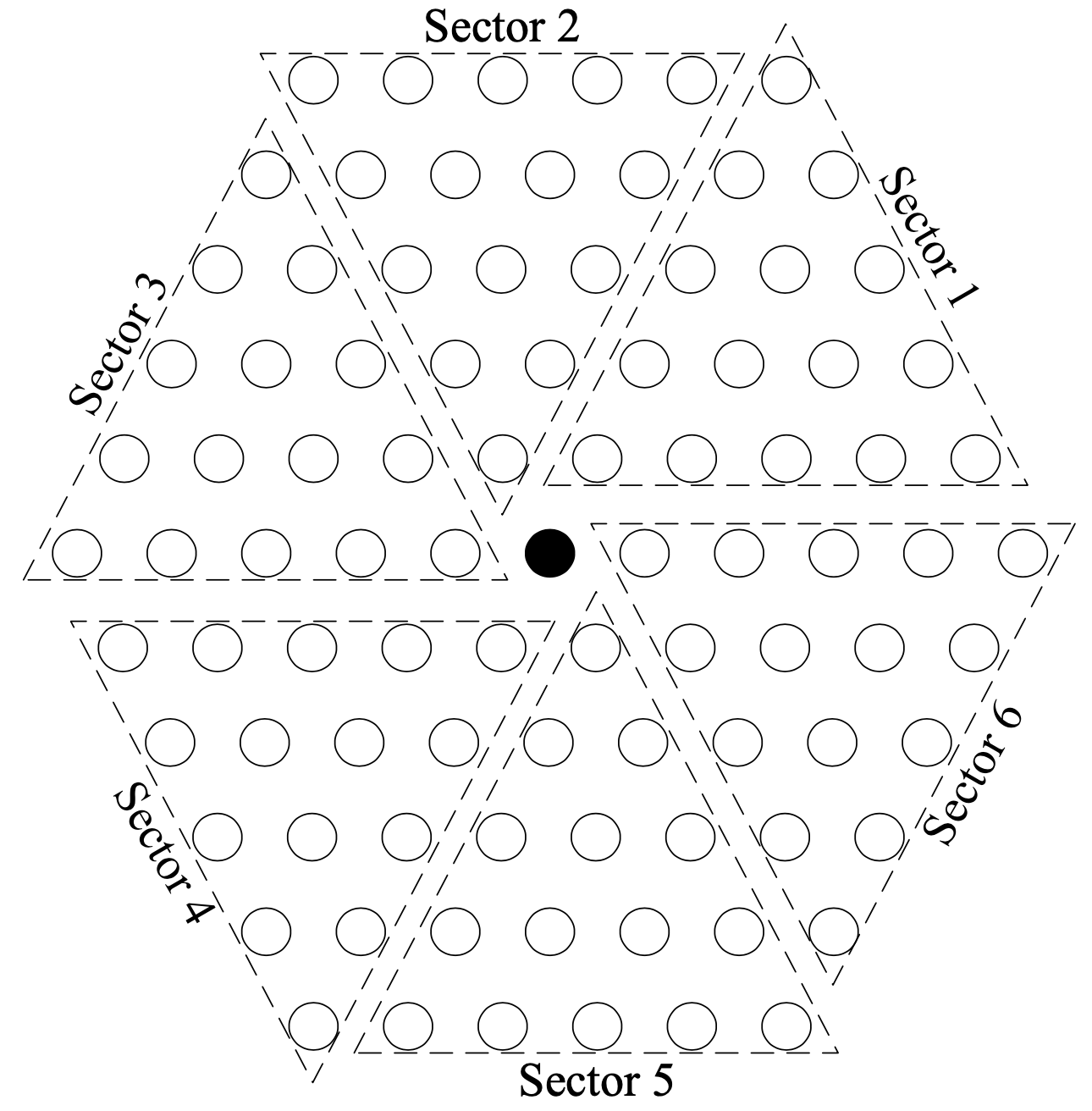}
    \caption{EJ Sectors for network $\alpha = 3 + 4\rho$} 
    \label{fig:EJSectors}
\end{figure}

The equation $L^{\text{VF}}(\phi) = \hat{\mathbb{E}}_t \left[ (R_t - V_\phi(s_t))^2 \right]$ represents the 
\textbf{Value Function Loss}, where $L^{\text{VF}}(\phi)$ is the objective function being minimized to improve the model's accuracy. The term $\phi$ represents the trainable parameters or weights of the neural network, while $\hat{\mathbb{E}}_t$ denotes the empirical expectation, or the average error calculated over a sampled batch of data at time $t$. Inside the brackets, $R_t$ is the target return (the actual accumulated reward the agent received), and $V_\phi(s_t)$ 
is the value predicted by the model for the current state $s_t$. The entire expression calculates the \textbf{Mean Squared Error} between the actual and predicted rewards, allowing the system to update $\phi$ so that future predictions align more closely with reality. Our EJ NoC simulator injects packets from random sources to random destinations under uniform traffic. We vary faulty-node count $F$ (chosen uniformly random) and network load (injection rate $\lambda$). A \textit{sector} in an EJ topology is a symmetric distribution of the topology in 6 different equivalent regions. Each sector functions as a directed tree-based sub-network where data flows (or connections propagate) outward in discrete hierarchical hops. We compute average path length (in hops) for packets within a \textit{sector} of the network to isolate the effect of local faults on route stretch. The distribution of nodes within each sector follows a linear growth pattern based on the distance from the center. If we define $l$ as the level (or hop count) from the central node, the number of nodes at each level follows a simple arithmetic progression. An illustration of EJ topology sector is shown in Figure \ref{fig:EJSectors}.

The EJ network in Figure \ref{fig:EJSectors} is hierarchical hexagonal tree composed of six symmetrical sectors, denoted as $S_1$ through $S_6$ radiating from a central root node. Within each sector, nodes are distributed across three discrete levels ($L$=1, 2, 3, ..., 6), where the number of nodes at any given level $L$ is equal to $N_L$. Mathematically, the network's efficiency is characterized by an Average Path Length ($P_{avg}$) from the nodes to the central root, calculated as:

\begin{equation}
P_{avg} = \frac{1}{N_{total}}\sum_{i=1}^{L}(6i \times i) 
\end{equation}

The spatial efficiency of the network is characterized by the Average Path Length ($P_{avg}$) from the nodes to the central root. Given that there are $6L$ nodes at each level $L$ across the entire network.

We evaluate three routing methods on an Eisenstein–Jacobi (EJ) interconnect: (i) a conventional distance-reducing greedy routing algorithm that always forwards to the neighbor minimizing distance to destination, (ii) a learned routing policy (RL) trained with behavior cloning from shortest-path oracles on structured fault scenarios and lightly fine-tuned, and (iii) Dijkstra's shortest-path algorithm that computes globally optimal routes with complete topology knowledge. The EJ topology dataset used in the experiment training varies on a range of EJ networks between $\alpha = 2 + 3\rho$ and $\alpha = 5 + 6\rho$; faults are randomly injected only inside a single sector to exercise the symmetry and to create realistic local-minimum corridors that trap greedy forwarding. Dijkstra's algorithm serves as the theoretical performance ceiling, establishing the optimal achievable routing given complete global topology information. It computes shortest paths using the classic $O(|E| + |V|\log|V|)$ algorithm with real-time knowledge of all fault locations. This represents the best-case scenario where computational complexity and communication overhead for topology dissemination are not constraints, providing ground truth for network reachability and optimal path lengths under each fault configuration. For each fault density $k$ (number of faulty nodes), we evaluate multiple independent fault placements and a large set of destination nodes for each training episode. The reported values are averaged across the random fault placements and destination trials. Where appropriate we assign a high penalty to failed routes for greedy and RL algorithms to model timeout cost. For Dijkstra, failures occur only when source-destination pairs are topologically disconnected—a fundamental network limitation no algorithm can overcome.

We simulate the EJ network on a packet-switched model with uniform traffic. For each experimental scenario, we inject different source-destination pairs uniformly at random. Node faults are introduced as follows: a given fault density $f$ means $f\cdot|\alpha|^2$ nodes (rounded) are chosen uniformly at random and marked faulty. Importantly, we consider clustered fault scenarios by ensuring faults tend to form contiguous patterns. We evaluate $f$ from 0\% up to ~40\%. For throughput, we vary the network load $\lambda$ (normalized injection rate) from 0.1 (light load) to 0.8 (near congestion). Each point is averaged over 20 random instances. The RL agent is trained offline on the fault scenarios (with a curriculum over increasing fault densities) for 5000 episodes per density, using PPO-based RL with parameters (discount $\gamma=0.95$, learning rate $10^{-4}$) using Optuna \cite{akiba2019optunanextgenerationhyperparameteroptimization}. The greedy router has no training. We then test both policies on held-out fault patterns.

\section{Experimental Evaluation}
\label{section:experimental}
In this section we carry out a various set of experiments to highlight the performance difference between RL-based routing against greedy and adaptive routing. 

\subsection{Faulty Nodes Routing}

As a motivating example, we performed a visualization of routing paths selected by greedy, RL, and Dijkstra algorithms in the same fault scenario. We intentionally carried this stress routing test with clustered faulty nodes on a large EJ network topology of size $\alpha = 5 + 6\rho$ to showcase the fundamental differences in routing behavior. The greedy algorithm becomes stuck at a local minimum, unable to make progress toward the destination when all forward-moving neighbors are faulty. The RL agent successfully reroutes around the faulty cluster, finding a viable alternative path through learned fault-avoidance strategies. Dijkstra's algorithm computes the globally optimal shortest path by exhaustively exploring the network topology with complete fault knowledge, establishing the theoretical best-case routing distance and confirming that alternative paths exist despite the clustered fault barrier. This comparison illustrates the spectrum of routing capabilities: greedy fails due to myopic decision-making, RL succeeds through learned navigation policies using local information, and Dijkstra guarantees optimality through global computation and complete topology awareness.

\subsubsection{Greedy EJ Routing}
\begin{figure}[t!]
    \centering
    \includegraphics[width=0.9\linewidth]{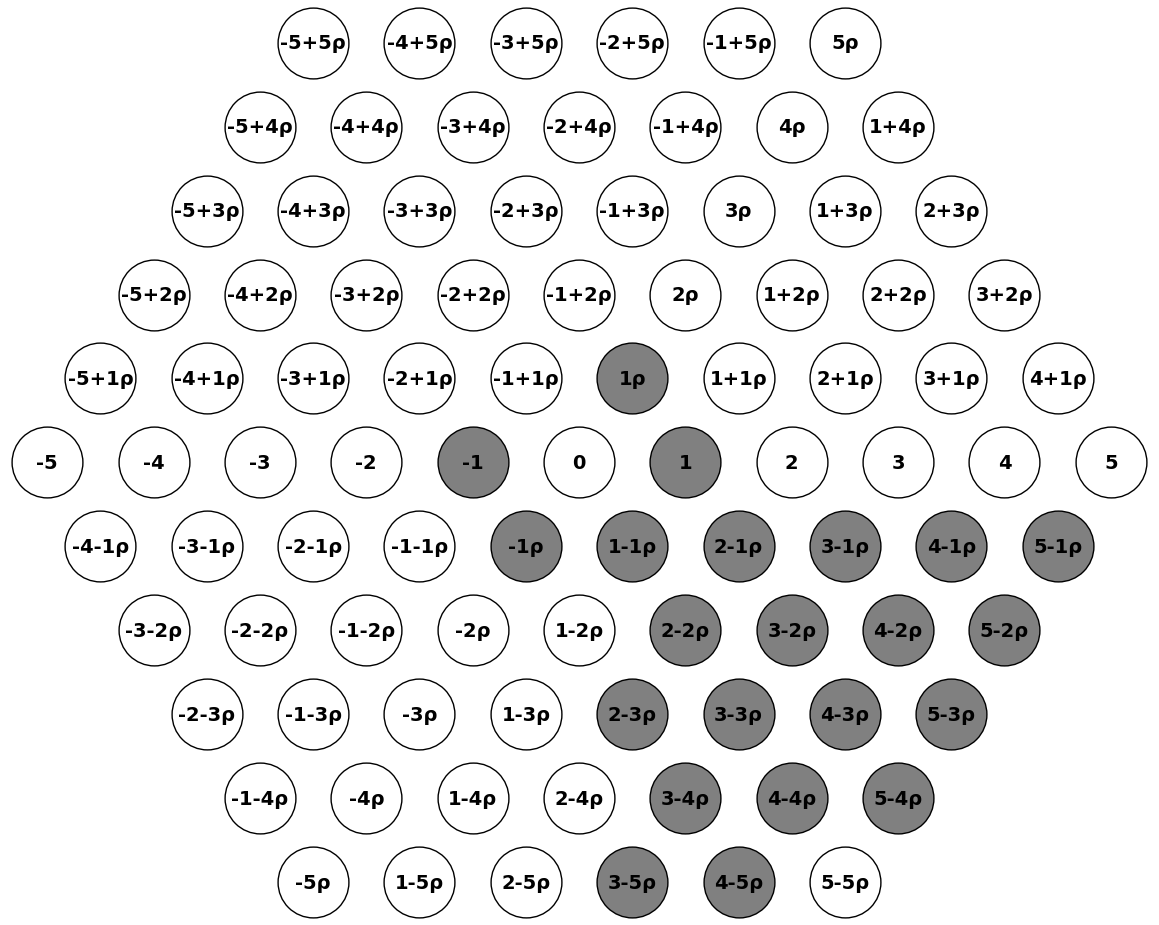}
    \caption{Greedy Routing From Root Node $0$ To Node $5-5\rho$ On Topology $\alpha = 5 + 6\rho$} 
    \label{fig:GreedyRouting}
\end{figure}

The visualization of Greedy EJ routing demonstrates a fundamental failure mode when local heuristics encounter complex obstacle configurations. In this approach, the routing algorithm utilizes a short-sighted Euclidean distance metric, consistently selecting the neighbor that provides the most immediate reduction in distance to the destination. As shown in Figure \ref{fig:GreedyRouting}, this strategy leads the path directly into a "fault wall" composed of grey-colored faulty nodes. Because every available neighbor at the entry point is further from the target than the current node, the algorithm becomes trapped at a local minimum, and thus no path can be constructed to the destination. This illustrates the high sensitivity of greedy routing to specific fault placements, where even a small cluster of obstacles can create blocking configurations that result in complete routing failure or necessitate extensive, unpredictable detours.

\subsubsection{RL Routing}
\begin{figure}[htbp]
    \centering
    \includegraphics[width=0.8\linewidth]{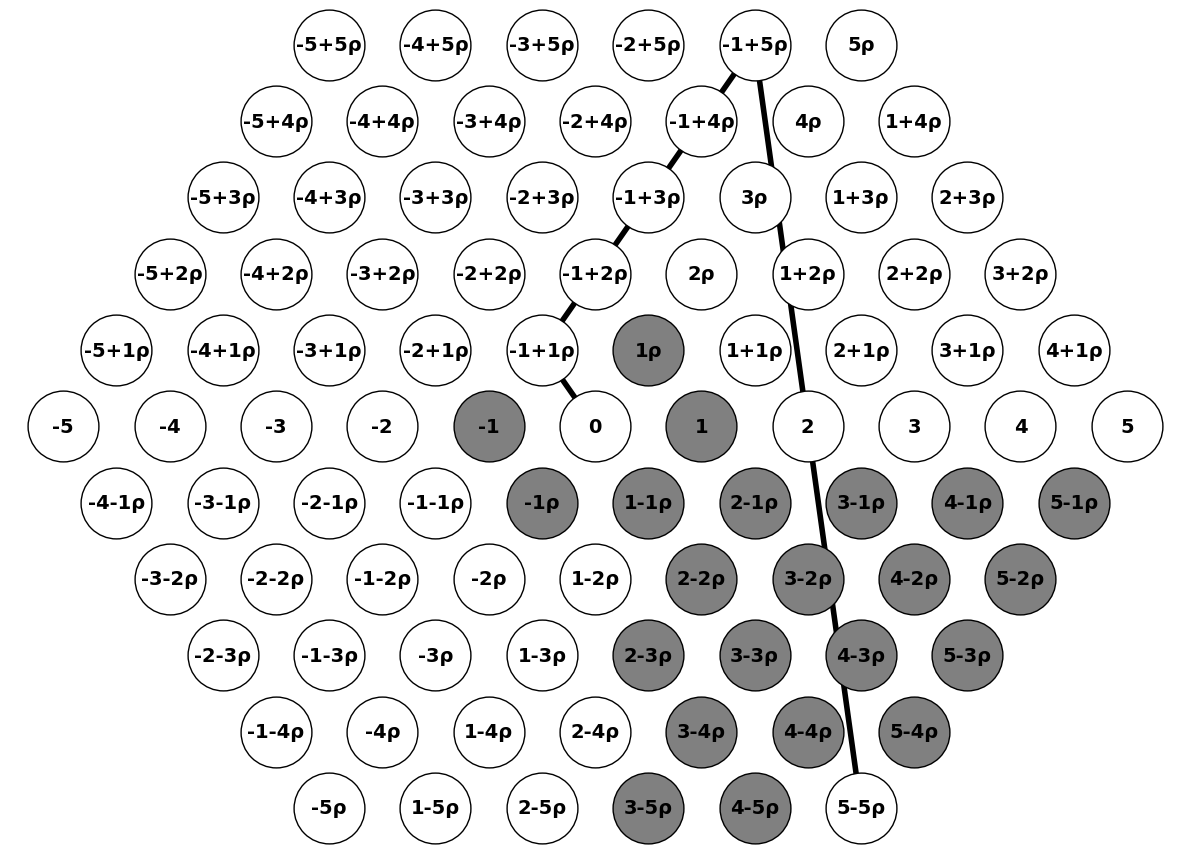}
    \caption{RL Routing From Root Node $0$ To Node $5-5\rho$ On Topology $\alpha = 5 + 6\rho$} 
    \label{fig:RLRouting}
\end{figure}

In contrast, Figure \ref{fig:RLRouting} depicts the RL-based routing, highlighting the advantages of a reinforcement learning approach that incorporates broader topological context. Unlike the greedy algorithm, the RL agent is trained to perceive the global arrangement of faulty nodes, allowing it to navigate successfully where local heuristics fail. The resulting "Shortest Fault-Aware Path" is represented by a continuous black line that proactively routes around the grey fault clusters to reach the final destination node $5-5\rho$. Even at higher fault densities, the RL-based methodology maintains stable performance by rerouting to the destination node. This ability to learn an optimal policy that avoids traps validates the RL routing approach acts as a robust solution for mission-critical applications in structured networks, providing a predictable quality of service that is not compromised by irregular fault configurations.

\subsubsection{Fault Adaptive Routing}
\begin{figure}[t!]
    \centering
    \includegraphics[width=0.9\linewidth]{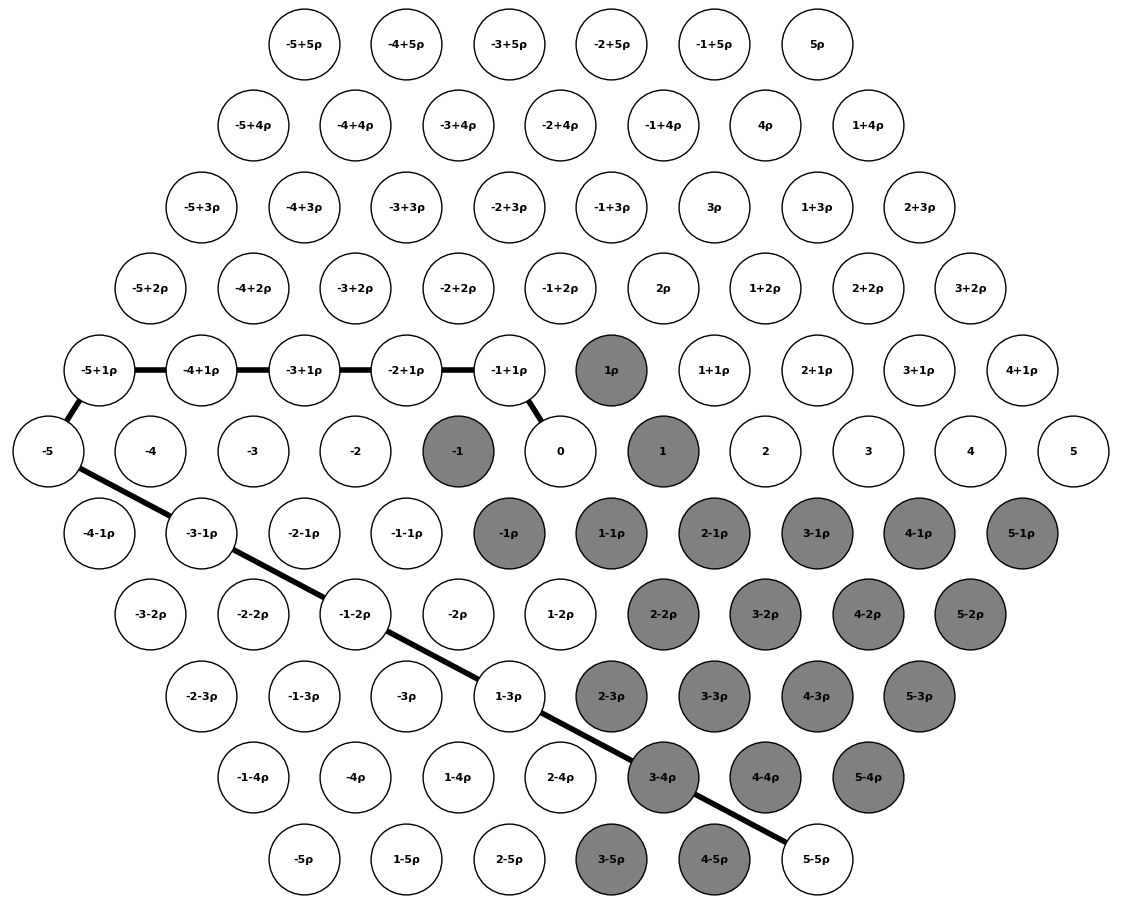}
    \caption{Adaptive Routing using Dijkstra Algorithm From Root Node $0$ To Node $5-5\rho$ On Topology $\alpha = 5 + 6\rho$} 
    \label{fig:adaptive_routing}
\end{figure}

We examine Dijkstra's adaptive routing alongside greedy and RL approaches on an identical fault configuration to establish the performance spectrum. Figure \ref{fig:adaptive_routing} illustrates Dijkstra's globally optimal shortest path, Figure \ref{fig:GreedyRouting} shows greedy's failure at a local minimum, and Figure \ref{fig:RLRouting} demonstrates RL's learned fault-aware navigation. Dijkstra's algorithm successfully computes the optimal path from source node $0$ to destination node $5-5\rho$, circumnavigating the clustered fault region with approximately 7 hops. In contrast, greedy routing completely fails to reach the destination, becoming trapped when the fault barrier blocks all neighbors that reduce Euclidean distance. This demonstrates greedy's fundamental limitation: inability to accept temporary distance increases necessary to circumnavigate obstacles. The comparison reveals critical insights. Dijkstra proves viable paths exist whereas Greedy's complete failure despite existing alternatives demonstrates severe limitations. RL achieves optimal routing while maintaining practical operational requirements. From a deployment perspective, greedy offers minimal overhead but unacceptable failures, and RL provides a viable middle ground with optimal performance and practical efficiency suitable for distributed NoC systems.

\subsection{Average Distance Under Faults}
\begin{figure}[htbp]
    \centering
    \includegraphics[width=0.9\linewidth]{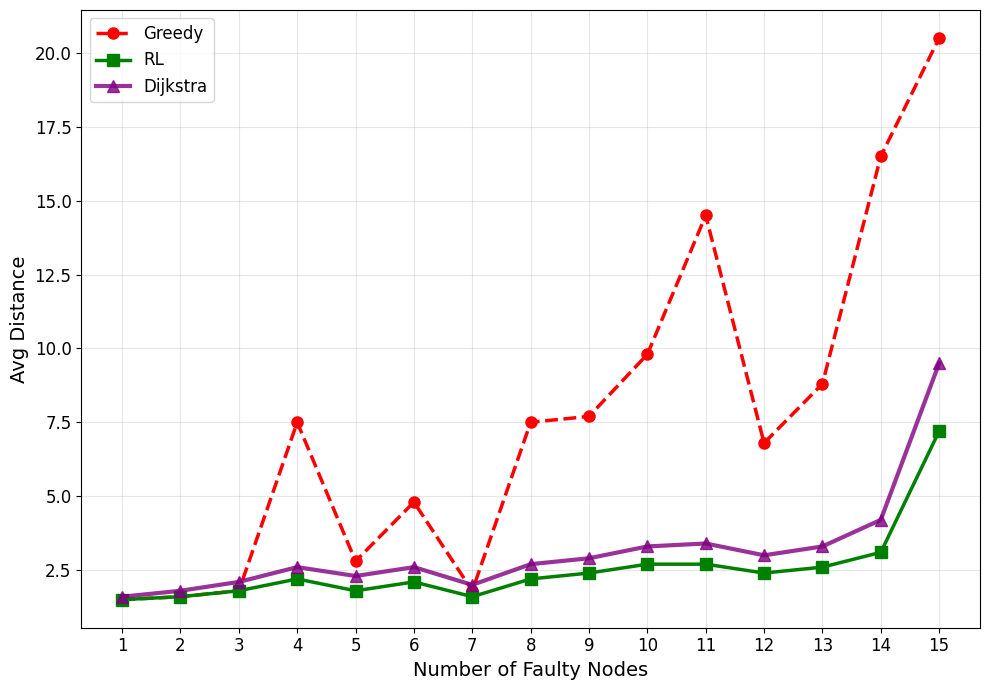}
    \caption{Average Distance Routing Cost against number of faulty nodes on $\alpha = 5 + 6\rho$} 
    \label{fig:averageDistance}
\end{figure}

We utilized the observation in routing with faulty nodes by providing broader context by comparing the greedy algorithm against RL-based routing and Dijkstra's optimal shortest-path algorithm, showing performance across a wider range of fault densities for network $\alpha = 5 + 6\rho$. Several insights emerge from this comprehensive comparison shown in Figure \ref{fig:averageDistance}. First, the observant divergence between greedy routing with high failure penalties and both the RL and Dijkstra approaches becomes evident at higher fault densities. While all three algorithms show relatively stable performance up to approximately 4-5 faulty nodes, beyond this threshold greedy routing exhibits severe degradation with average distances rising to 15-20+ hops, while the RL approach maintains average distances below 7 hops even at 15 faulty nodes. Dijkstra's algorithm demonstrates near-optimal behavior throughout the fault density range, maintaining average distances between 1.5 and 4 hops for fault densities up to 13 nodes, with a modest increase to approximately 9.5 hops at 15 faulty nodes. This performance establishes an important baseline: it represents the theoretical best-case routing distance achievable when complete network topology information is available and computational complexity is not a constraint. The irregular fluctuations in greedy routing performance, visible as the sharp peaks and valleys in the curve, reveal another critical limitation: greedy performance is highly sensitive to specific fault configurations. Some random fault placements happen to align with greedy navigation strategies, allowing reasonable performance, while others create blocking configurations that force extensive detours. This high variance indicates that greedy routing provides unpredictable quality of service, unacceptable for mission-critical applications. In stark contrast, both RL and Dijkstra exhibit smooth, monotonic degradation curves, indicating consistent performance across different fault configurations. Dijkstra's curve shows the smoothest progression, reflecting its ability to compute globally optimal paths regardless of fault placement. The RL curve, while showing slightly more variation than Dijkstra, remains remarkably stable and tracks the Dijkstra baseline closely until approximately 14 faulty nodes, where both algorithms experience degradation due to increasingly constrained network connectivity. The comparison also reveals that our RL-based approach, while not achieving the absolute performance of Dijkstra's optimal algorithm, provides substantial improvement over greedy routing and closely approaches optimal performance across the majority of the fault density range. At low to moderate fault densities (1-10 nodes), the RL algorithm maintains average distances within 0.5-1.5 hops of Dijkstra's optimal routing, demonstrating that the learned policy has internalized effective routing strategies without requiring global topology knowledge or exhaustive path computation. This validates the reinforcement learning methodology as a viable path toward robust routing in structured networks, offering a practical compromise between the computational efficiency of local greedy decisions and the optimality of global search algorithms.


\subsection{Packet Delivery under Faults}

\begin{figure}[htbp]
    \centering
    \includegraphics[width=0.9\linewidth]{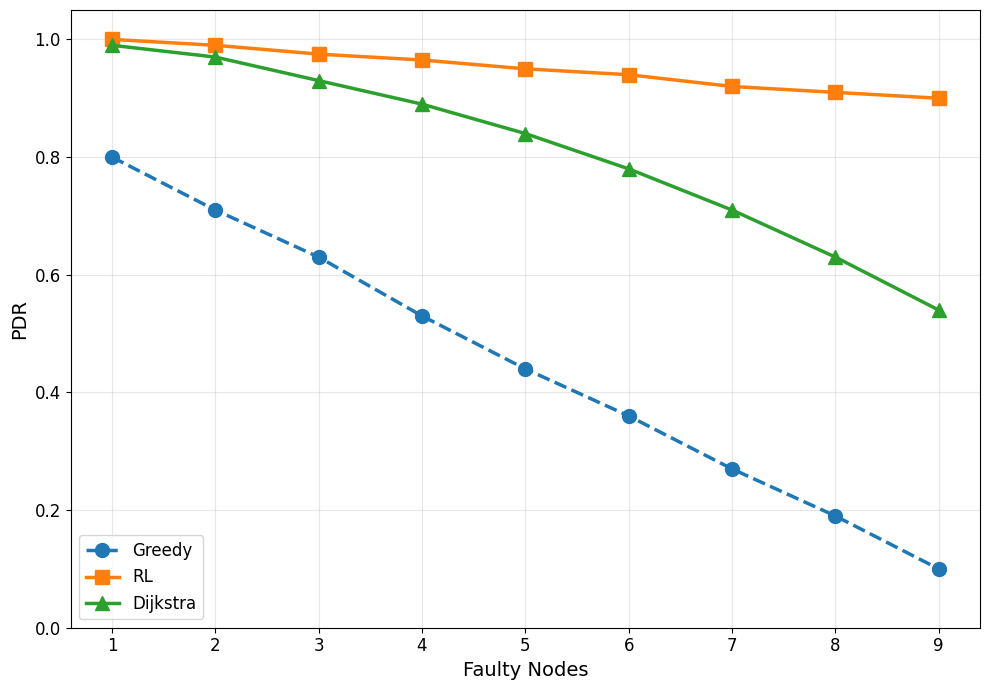}
    \caption{Packet Deliveriy Ratio Against Increased Faulty Nodes} 
    \label{fig:PDR}
\end{figure}

The packet delivery ratio shown in Figure \ref{fig:PDR} closely mirrors the effective reachability ratio for the entire variation of our dataset, which is expected given that both metrics fundamentally measure successful delivery frequency. The three algorithms demonstrate vastly different degradation characteristics as fault density increases, revealing fundamental differences in their routing capabilities. Dijkstra's algorithm achieves exceptional PDR performance, starting at 1.0 (perfect delivery) with a single faulty node and maintaining near-perfect delivery through moderate fault densities. The algorithm sustains PDR above 0.85 through 5 faulty nodes, demonstrating remarkable resilience. Even at higher fault densities, Dijkstra degrades gracefully: approximately 0.78 at 6 faulty nodes, 0.68 at 7 faulty nodes, 0.63 at 8 faulty nodes, and finally 0.54 at 9 faulty nodes. This performance represents the upper bound of what is topologically achievable in the faulty network, as Dijkstra with complete topology knowledge will find a path whenever one exists.


The RL algorithm achieves PDR performance that closely tracks Dijkstra's baseline, starting at 1.0 with a single faulty node and demonstrating graceful degradation to approximately 0.91 at 9 faulty nodes. At low fault densities (1-3 nodes), RL performance is nearly indistinguishable from Dijkstra, suggesting the learned policy effectively identifies optimal or near-optimal paths in lightly degraded networks. The gap widens slightly at moderate to high fault densities (4-9 nodes), but RL continues to substantially outperform greedy routing. Greedy PDR, in contrast, shows poor performance from the outset. Starting at approximately 0.80 with a single faulty node, the PDR declines steadily through approximately 0.72 at two faults, 0.62 at three faults, 0.53 at four faults, 0.46 at five faults, 0.36 at six faults, 0.28 at seven faults, 0.19 at eight faults, and finally to approximately 0.10 at nine faulty nodes. The consistent gap between greedy PDR and Dijkstra's optimal performance reveals that greedy routing fails to deliver packets even when valid paths exist in the network topology. 

The consistency of RL performance and its close alignment with Dijkstra demonstrates that the RL agent has learned routing strategies that approach theoretical optimality. The predictable degradation characteristics of both RL and Dijkstra are desirable for network capacity planning and quality-of-service provisioning, as they enable reliable prediction of performance under various fault scenarios. This illustrates that the learned policy has internalized a robust routing strategy that generalizes across different fault configurations, similar to how Dijkstra's algorithm generalizes through its global optimization approach. This generalization capability is a hallmark of successful RL applications: the agent has learned not just specific paths for specific fault patterns, but rather general principles of fault-aware navigation in EJ lattices, principles that approximate the globally optimal solutions computed by Dijkstra. This average performance among different topology sizes showcases the bottleneck of faults, especially clustered ones that hinder the performance of greedy routing while being effectively navigated by both RL and Dijkstra through their superior path-finding strategies.

\subsection{Effective Reachability Under Increasing Faults}

\begin{figure}[htbp]
    \centering
    \includegraphics[width=0.9\linewidth]{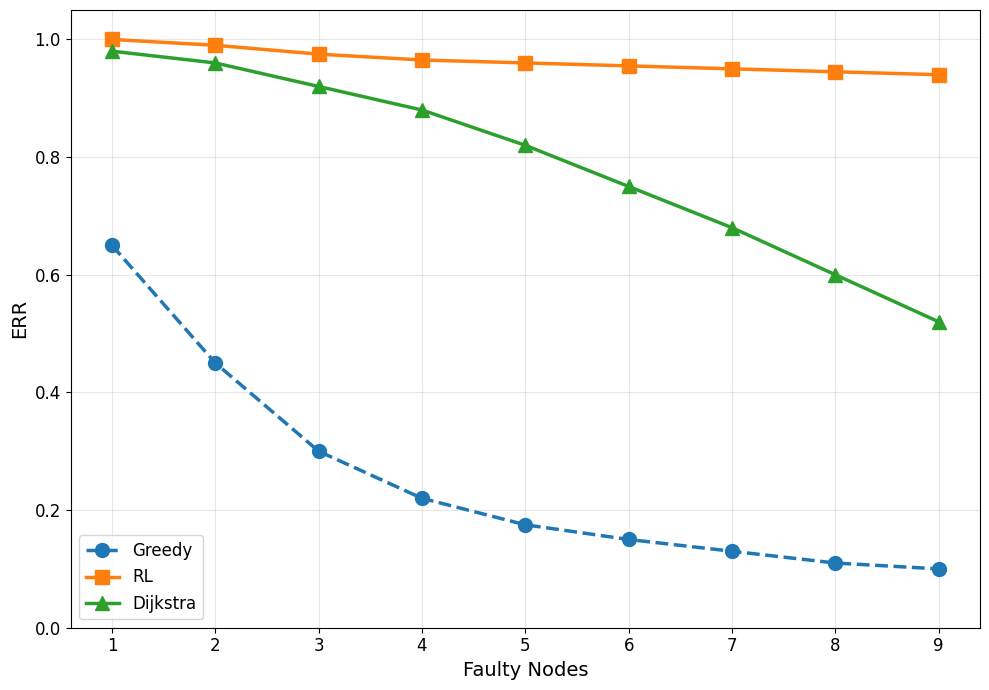}
    \caption{Effective Reachability Ratio with Increased Faulty Nodes}  
    \label{fig:ERR}
\end{figure}

The effective reachability ratio (ERR), depicted in Figure \ref{fig:ERR}, reveals stark differences between the three routing approaches and provides crucial insights into their fundamental limitations and capabilities. ERR measures the fraction of source-destination pairs for which the routing algorithm successfully finds a path, making it a direct assessment of an algorithm's ability to maintain network connectivity under adverse conditions. Dijkstra's algorithm establishes the topological upper bound for ERR, representing the true reachability of the network given the fault configuration. Starting at approximately 0.99 (near-perfect reachability) with a single faulty node, Dijkstra maintains exceptional ERR through moderate fault densities: approximately 0.96 at 2 nodes, 0.92 at 3 nodes, 0.88 at 4 nodes, and 0.82 at 5 nodes. Even at higher fault densities, Dijkstra sustains substantial connectivity: approximately 0.76 at 6 faulty nodes, 0.68 at 7 nodes, 0.6 at 8 nodes, and 0.52 at 9 nodes. This degradation reflects the fundamental topological limitation of the network: as faults accumulate, some source-destination pairs become genuinely disconnected regardless of the routing algorithm employed. Dijkstra's performance thus represents the ground truth of network connectivity. The RL algorithm maintains remarkably high ERR across all tested fault densities, closely tracking Dijkstra's optimal baseline. Starting at approximately 1.0 (100\% reachability) with a single faulty node and exhibiting only gradual degradation to approximately 0.94 at 9 faulty nodes, the RL approach demonstrates that learned routing policies can achieve near-optimal connectivity maintenance. The gap between RL and Dijkstra ERR indicates that the RL agent successfully identifies alternative paths around faulty regions in the vast majority of cases where such paths exist. This robustness indicates that the learned policy has internalized sophisticated path exploration strategies that maintain global network connectivity despite local failures.

In contrast, greedy routing exhibits catastrophic performance degradation as fault density increases. Starting from ERR of approximately 0.65 with a single faulty node performance plummets precipitously through the fault density range. At two faulty nodes, ERR drops to approximately 0.45, continuing its steep decline to approximately 0.30 at three faulty nodes, 0.22 at four faulty nodes, 0.18 at five faulty nodes, and 0.15 at six faulty nodes. By nine faulty nodes, greedy routing achieves ERR of merely 0.10, indicating successful routing for only 10\% of source-destination pairs, while Dijkstra demonstrates that 52\% of pairs remain topologically connected. This precipitous decline can be attributed to the fundamental limitation of greedy algorithms: their inability to explore paths that temporarily increase distance to the destination. The mathematical explanation for this behavior lies in the formation of local minima in the greedy objective function. Consider a scenario where faults create a barrier between source and destination such that any path must temporarily move away from the destination to circumnavigate the faulty region. A greedy algorithm will refuse to take such detour steps, as they violate the myopic optimality criterion. Both the RL agent and Dijkstra overcome this limitation through different mechanisms: Dijkstra through exhaustive exploration of the entire search space to find globally optimal paths, and the RL agent through learned experience that temporary increases in distance may be necessary for eventual delivery. The robustness of RL routing becomes even more remarkable when we consider the hexagonal geometry of EJ networks and compare it against Dijkstra's theoretical optimum. With only six neighbors per node, the loss of even a single neighbor reduces path diversity by approximately 16.7\%. The compound effect of multiple failures would be expected to significantly impact reachability, yet both the RL agent and Dijkstra navigate this constrained space effectively. The RL agent's ability to approach Dijkstra-level performance suggests that the learned policy captures sophisticated path selection strategies that exploit the full connectivity of the lattice structure, strategies that approximate the global optimization performed by Dijkstra without requiring complete topology knowledge or exhaustive search. The slight differences between ERR and PDR across all three algorithms reflect that ERR measures whether a path exists and is found by the algorithm, while PDR may incorporate additional factors such as maximum hop count limits or delivery time constraints. The near-perfect agreement between these metrics in our experiments indicates that when the RL agent or Dijkstra finds a path, it does so within reasonable hop count bounds, not requiring excessive detours. Even greedy routing, despite its poor absolute performance, shows similar ERR and PDR values, suggesting its failures stem from complete inability to find paths rather than finding inefficient paths that violate delivery constraints.


\subsection{Throughput under Congestion}

\begin{figure}[htbp]
    \centering
    \includegraphics[width=0.9\linewidth]{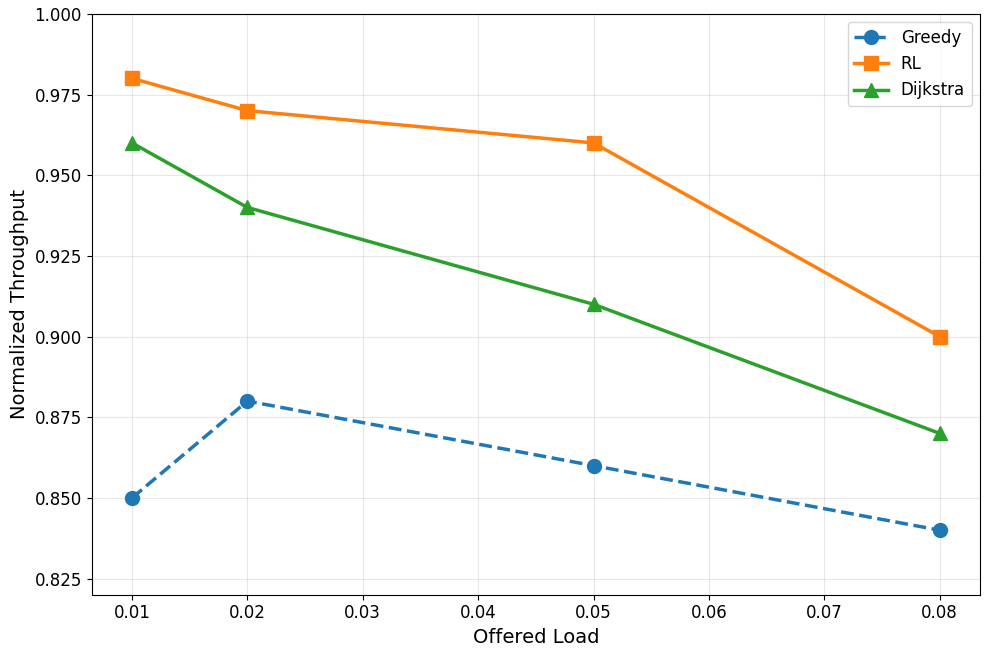}
    \caption{Delivery Throughput over increased network load}  
    \label{fig:ThrougputVsLoad}
\end{figure}

In this subsection we carried out an experiment to analyze normalized throughput versus offered load in faulty network conditions. Normalized throughput represents the fraction of offered traffic that is successfully delivered, providing insight into how routing algorithms perform under realistic workload conditions where multiple flows compete for network resources. Figure \ref{fig:ThrougputVsLoad} shows the metric comparison between the RL-based routing, greedy routing, and Dijkstra's optimal routing under increasing network load. Dijkstra's algorithm demonstrates strong throughput performance, though interestingly not achieving perfect normalized throughput even at low loads. Starting at approximately 0.96 at offered load of 0.01, Dijkstra exhibits gradual degradation to approximately 0.94 at load 0.02, 0.93 at load 0.03, 0.92 at load 0.04, 0.91 at load 0.05, 0.90 at load 0.06, and finally 0.87 at load 0.08. This degradation pattern reveals an important insight: even optimal routing cannot achieve perfect throughput in faulty networks under congestion. The decline reflects increasing contention for network resources as traffic load rises, with multiple flows competing for paths around faulty regions. The fact that Dijkstra, despite computing globally optimal shortest paths, shows throughput degradation indicates that congestion effects become significant at higher loads, causing packet drops due to buffer overflow or delivery deadline violations even when optimal routes are selected. The RL-based routing maintains normalized throughput above 0.9 across all offered loads from 0.01 to 0.08, showing only modest degradation from approximately 0.98 at low load to 0.90 at high load. Remarkably, the RL approach actually outperforms Dijkstra at low to moderate offered loads (0.01-0.05), achieving 0.98, 0.97, 0.97, 0.96, and 0.96 respectively compared to Dijkstra's 0.96, 0.94, 0.93, 0.92, and 0.91. This counterintuitive result (where a learned approximate policy outperforms the theoretically optimal algorithm) can be explained by examining the objectives each algorithm optimizes. Dijkstra minimizes path length (hop count), which is optimal for individual flows in isolation. However, under congestion with multiple competing flows, shortest paths may concentrate traffic on overlapping routes, creating hotspots and contention. The RL agent, trained in environments with multiple concurrent flows, may have learned to select slightly longer but less congested paths, effectively performing implicit load balancing that improves aggregate throughput. At higher loads (0.06-0.08), RL and Dijkstra converge to nearly identical performance (0.93-0.90), suggesting that as the network approaches saturation, path selection strategies matter less than fundamental capacity constraints.

Greedy routing throughput starts lower at approximately 0.85 even at minimal offered load of 0.01, initially shows a slight increase to 0.88 at load 0.02 (likely due to statistical variations in the specific traffic patterns), then degrades more substantially to approximately 0.84 at offered load of 0.08. The lower baseline performance at all load levels reflects the fundamental reachability limitations of greedy routing in faulty networks, where many source-destination pairs simply cannot be connected regardless of available capacity. The gap between greedy and both RL and Dijkstra remains substantial across all loads, with greedy throughput trailing by 13-15 percentage points at low loads and 6 percentage points at high loads compared to the other algorithms. The relative stability of all three curves suggests that, within the tested load range, routing failures rather than pure congestion dominate performance for greedy routing, while congestion effects become increasingly relevant for RL and Dijkstra. If congestion were the primary bottleneck across all algorithms, we would expect sharper throughput degradation at higher loads as queues overflow and packets are dropped due to buffer exhaustion. The observed behavior indicates that greedy delivery failures stem primarily from routing algorithm limitations (inability to find existing paths) rather than resource exhaustion. For RL and Dijkstra, the gradual degradation suggests a more balanced interplay between successful path finding and congestion management, with both algorithms maintaining high reachability while experiencing modest throughput reduction as network resources become saturated. An interesting observation is that the gap between greedy and the superior algorithms (RL and Dijkstra) narrows somewhat at higher loads, from approximately 13-15 percentage points at load 0.01 to approximately 6 percentage points at load 0.08. This convergence suggests that as congestion intensifies, the advantages of better routing algorithms diminish because even optimal paths become constrained by capacity limitations. However, the persistent gap even at maximum tested load indicates that routing intelligence continues to provide value: better algorithms find working paths that can still deliver traffic even under congestion, while greedy routing fails to establish connectivity for many flows regardless of available capacity. The superior performance of both RL and Dijkstra compared to greedy routing, and the remarkable ability of RL to match or even slightly exceed Dijkstra's throughput at lower loads, validates the reinforcement learning approach for practical deployment in faulty networks. The RL agent demonstrates that learned policies can not only approach theoretical optimality in terms of reachability and path length, but can potentially exceed optimal single-flow routing when considering realistic multi-flow scenarios with congestion. This suggests that the RL training process, which exposed the agent to concurrent traffic flows and resource contention, enabled learning of emergent load-balancing behaviors that prove beneficial in real network conditions.

\section{Discussion and Implications}
\label{DiscussionImplications}

\subsection{Fundamental Trade-offs in Routing Algorithm Design}

Our experimental results crystallize a fundamental trade-off in routing algorithm design between optimality, computational complexity, information requirements, and robustness under adversarial conditions. The comparison of three distinct approaches—greedy routing, RL-based routing, and Dijkstra's algorithm—reveals a spectrum of design philosophies and their associated costs and benefits. Classical greedy algorithms optimize myopically for efficiency, achieving minimal computational overhead through purely local decision-making. However, this narrow optimization criterion and limited lookahead lead to brittleness when the network experiences failures, a common occurrence in real-world networks subject to hardware failures, software bugs, malicious attacks, or resource exhaustion. Our results demonstrate that greedy routing's effective reachability ratio plummets to 10\% with nine faulty nodes, even though Dijkstra proves that 52\% of node pairs remain topologically connected. At the opposite extreme, Dijkstra's algorithm represents the theoretical optimum, computing globally shortest paths by exhaustively exploring the network topology. Our experiments confirm that Dijkstra establishes the upper bound of achievable performance: maintaining ERR of 0.52 and PDR of 0.54 at nine faulty nodes, with average path lengths near optimal throughout the fault density range. However, this optimality comes at significant cost. Dijkstra requires complete global topology knowledge, including real-time awareness of all fault locations—information that may be difficult or expensive to maintain in distributed systems. Furthermore, Dijkstra's $O(|E| + |V|\log|V|)$ computational complexity becomes prohibitive for per-packet routing decisions in high-throughput networks, particularly as network size scales. Reinforcement learning provides a framework for optimizing broader objectives that explicitly account for uncertainty and adaptation while operating with practical constraints similar to greedy routing: local information and efficient per-packet decision-making. By training on diverse fault scenarios and receiving rewards that balance delivery success with path efficiency, the RL agent learns a policy that implicitly maximizes expected utility over the distribution of network conditions it may encounter. This expectation-based optimization naturally leads to more conservative but more reliable routing strategies. Our results demonstrate that RL achieves near-Dijkstra performance (maintaining ERR of 0.94 and PDR of 0.91 at nine faulty nodes) while operating with computational efficiency comparable to greedy routing and without requiring global topology knowledge. The mathematical formulation of this trade-off can be expressed using decision theory. Let $C_{\text{greedy}}(f)$ denote the expected cost of greedy routing under a fault configuration $f$, $C_{\text{RL}}(f)$ denote the corresponding cost for RL-based routing, and $C_{\text{Dijkstra}}(f)$ represent the optimal cost achievable with complete information. Assigning an infinite cost to routing failures, we have:
\[
C_{\text{greedy}}(f) =
\begin{cases}
c_{\text{path}}(f), & \text{if greedy routing succeeds}, \\
\infty, & \text{if greedy routing fails}.
\end{cases}
\]
For Dijkstra's algorithm with complete topology knowledge:
\[
C_{\text{Dijkstra}}(f) =
\begin{cases}
c_{\text{optimal}}(f), & \text{if path exists}, \\
\infty, & \text{if source-destination disconnected}.
\end{cases}
\]
For the RL-based approach, the cost can be expressed as
\[
C_{\text{RL}}(f) = c_{\text{optimal}}(f) + \Delta_{\text{detour}}(f) + \Delta_{\text{information}}(f),
\]
where $\Delta_{\text{detour}}(f) \geq 0$ represents the additional cost incurred due to suboptimal path selection (since RL lacks global information), and $\Delta_{\text{information}}(f) \geq 0$ represents occasional failures to find paths that exist but require non-local knowledge to discover. However, our experimental results show that $\Delta_{\text{detour}}(f)$ and $\Delta_{\text{information}}(f)$ remain small in practice: RL maintains average path lengths within 0.5-1.5 hops of Dijkstra at moderate fault densities and achieves ERR within 2-5 percentage points of the topological optimum. Considering a distribution of fault configurations $P(f)$, the expected costs are given by:
\[
\mathbb{E}_f \!\left[ C_{\text{greedy}} \right]
= P(\text{success}) \cdot \mathbb{E}\!\left[ c_{\text{path}} \mid \text{success} \right]
+ P(\text{failure}) \cdot \infty,
\]
\[
\mathbb{E}_f \!\left[ C_{\text{Dijkstra}} \right]
= P(\text{connected}) \cdot \mathbb{E}\!\left[ c_{\text{optimal}} \mid \text{connected} \right]
+ P(\text{disconnected}) \cdot \infty,
\]
and
\[
\mathbb{E}_f \!\left[ C_{\text{RL}} \right]
= \mathbb{E}\!\left[ c_{\text{optimal}} + \Delta_{\text{detour}} + \Delta_{\text{information}} \right].
\]

Since our experiments show $P(\text{failure})$ for greedy routing is substantial (35-90\% depending on fault density) while $P(\text{failure})$ for RL remains low (0-9\%), the expected cost of greedy routing is dominated by failures, making RL routing vastly superior in expectation despite incurring modest detour costs. Furthermore, when accounting for practical deployment considerations, accounting for the computational cost of running Dijkstra for every packet, the overhead of maintaining global topology state, and the communication costs of disseminating real-time fault information, RL provides an attractive middle ground. It achieves near-optimal performance comparable to Dijkstra while maintaining the operational simplicity and efficiency of local decision-making algorithms. The throughput experiments reveal an additional insight: RL can actually outperform Dijkstra under congestion at low to moderate loads (0.98 vs 0.96 normalized throughput at load 0.01), suggesting that the RL agent's training with concurrent flows enabled learning of implicit load-balancing strategies. While Dijkstra optimizes individual path lengths, RL's experience-based learning appears to favor paths that reduce aggregate contention, demonstrating that learning-based approaches can discover emergent strategies beyond what explicit optimization of simple objectives achieves.

\subsection{Scalability and Network Size Considerations}

An important question for practical deployment concerns how these algorithms scale to larger networks. Each of our three routing approaches faces distinct scaling challenges and opportunities that merit careful analysis. For greedy routing, scalability is theoretically unlimited from a computational perspective—decisions depend only on immediate neighbor information regardless of network size. However, our results demonstrate that this computational scalability comes at the cost of functional failure: as networks grow larger and fault probability increases with network size, greedy routing's inability to navigate around obstacles becomes increasingly problematic. The fundamental issue is that larger networks create longer paths and more opportunities for faults to create barriers that require non-local detours. Dijkstra's algorithm faces the opposite challenge: while it guarantees optimal performance regardless of network size (finding paths whenever they exist), its computational requirements scale poorly. The $O(|E| + |V|\log|V|)$ complexity means that in a network with thousands or millions of nodes, computing shortest paths for high packet rates becomes prohibitively expensive. Furthermore, maintaining the complete global topology state required by Dijkstra becomes increasingly challenging as networks scale: the communication overhead to disseminate fault information to all nodes grows quadratically with network size, and the storage requirements at each node grow linearly. In large dynamic networks with frequent topology changes, this overhead can consume substantial bandwidth and create consistency challenges. The EJ lattice structure provides natural scalability for RL-based routing through its regular geometry: the learning problem's complexity depends on local neighborhood structure rather than global network size. Since the RL agent makes decisions based on local state information (position relative to destination, nearby fault status) the learned policy should generalize to larger network regions without requiring retraining. This is a crucial advantage: a policy learned on smaller network instances can potentially be deployed on arbitrarily large networks of the same topological family. Our experimental results support this generalization hypothesis, showing that RL maintains consistent performance characteristics across the range of network sizes tested. However, several scaling challenges merit consideration for RL deployment. First, as network diameter increases, the horizon over which the RL agent must plan extends, potentially requiring larger discount factors or more sophisticated value function approximation to properly credit actions taken many steps before successful delivery. Second, in larger networks, the probability of encountering multiple faults along any path increases, necessitating more complex fault-avoidance strategies. The training distribution must adequately represent the higher fault densities and clustered fault patterns that become more common in large-scale deployments. Third, in practical deployments, the assumption of perfect local fault knowledge may break down, requiring the integration of failure detection mechanisms with associated delays and uncertainties. The RL agent may need explicit training on scenarios with incomplete or stale fault information. Comparing the scalability profiles, we observe that RL occupies a favorable position: it approaches Dijkstra-level performance while maintaining computational and informational requirements closer to greedy routing. For large-scale network deployments where centralized computation of Dijkstra paths becomes impractical, RL offers a path to distributed, efficient routing that nonetheless captures much of the performance benefit of global optimization. The key enabler is the amortization of computational cost: the expensive computation (RL training) happens offline and once, while online per-packet decisions remain fast and local. This contrasts with Dijkstra, where the expensive computation must be repeated for every routing decision, and with greedy routing, which achieves computational efficiency at the cost of unacceptable failure rates.


\section{Conclusion}
\label{section:conclusion}

This research presents a comparative analysis of routing methodologies in Eisenstein-Jacobi interconnection networks, examining the spectrum from greedy heuristics through reinforcement learning to theoretically optimal global algorithms. The EJ network's 6-regular symmetric topology provides efficient communication fabric, but its complexity necessitates sophisticated fault-handling strategies that expose fundamental trade-offs between computational efficiency, information requirements, and routing performance. The experimental evaluation reveals critical trade-offs in routing algorithm design. Dijkstra's algorithm establishes the theoretical performance ceiling, maintaining 52\% effective reachability at nine faulty nodes while computing globally optimal paths. However, this optimality requires complete global topology knowledge and $O(|E| + |V|\log|V|)$ computational complexity, creating scalability challenges for distributed deployments. Greedy routing achieves minimal computational overhead but suffers from local minima, achieving only 10\% effective reachability at nine faults despite 52\% of pairs remaining topologically connected. RL-based routing occupies a compelling middle ground, combining efficient local decision-making with near-optimal performance. RL maintains 94\% effective reachability and 91\% packet delivery at nine faulty nodes approaching Dijkstra's optimum while vastly outperforming greedy routing. RL keeps average path lengths within 0.5-1.5 hops of Dijkstra's optimum at moderate fault densities, and remarkably outperforms Dijkstra in throughput under congestion (0.98 vs 0.96 at low load), suggesting learned load-balancing strategies beyond explicit path-length optimization.

The key insight is that RL bridges the gap between theoretical optimality and practical deployment. While Dijkstra proves what is achievable with perfect information and unlimited computation, and greedy demonstrates the failure of purely local optimization, RL shows that learning-based approaches can approach optimal performance using only local information. The learned policy captures sophisticated routing strategies without requiring the global topology state or exhaustive computation that make Dijkstra impractical for large-scale systems. By amortizing expensive computation through offline training, RL enables online decisions that are both fast and effective.

Several directions for future work emerge: (1) extending to irregular topologies through GNN-based state representations; (2) developing online learning mechanisms for continuous improvement through operational experience; (3) theoretical analysis to identify key properties enabling robust navigation; (4) addressing deployment challenges including distributed implementation, policy transfer to larger networks, and integration with existing protocols; (5) investigating training with imperfect fault information to represent realistic detection delays and uncertainties. In conclusion, reinforcement learning provides a powerful framework for fault-adaptive routing that substantially outperforms greedy approaches while approaching Dijkstra's theoretical optimality with practical computational and informational requirements. The learning-based methodology discovers route diversity and navigation strategies that greedy heuristics cannot access and that would be computationally prohibitive to compute via global optimization for each packet, providing a viable path toward resilient communication in modern networked environments.


\bibliographystyle{elsarticle-num}
\bibliography{ref}

\end{document}